\documentclass[11pt]{llncs}
\usepackage{url,upref,amscd,amssymb,amsmath}
\begin{document}
\title{A simple generalization of the El-Gamal cryptosystem to non-abelian
groups}
\author{Ayan Mahalanobis}
\institute{Department of Mathematical Sciences, Stevens Institute of
  Technology, Hoboken, NJ 07030.\\
\today\\
\email{Ayan.Mahalanobis@stevens.edu}}
\maketitle
\begin{abstract}In this paper we study the MOR cryptosystem. We use
  the group of unitriangular matrices over a finite field as the
  non-abelian group in the MOR cryptosystem. We show that a
  cryptosystem similar to the El-Gamal cryptosystem over finite fields
  can be built using the proposed groups and a set of automorphisms of
  these groups. We also show that the security of this proposed MOR
  cryptosystem is equivalent to the El-Gamal cryptosystem over finite fields.
\end{abstract}
\noindent\emph{Keywords:} MOR Cryptosystem, Unitriangular Matrices.
\section{Introduction}
Most of the public key cryptosystems popular today are built on abelian
groups. It is natural to try to generalize these cryptosystems to non-abelian groups,
not only because the current systems are getting old with time, but
also there is an interesting academic adventure in trying to do
so. The cryptosystem that we have in mind is the \emph{El-Gamal}
cryptosystem \cite[Section 2]{menezes-koblitz}
which is built on the \emph{Discrete Logarithm Problem} \cite[Section
2]{menezes-koblitz}. The discrete 
logarithm problem can be generalized in different ways, to mention
just two of them -- one was
done in \cite{thesis} and the other is the MOR cryptosystem
\cite{crypto2001}.

The MOR cryptosystem has attracted a lot of attention and some well
written papers \cite{asiacrypt2004,paeng,tobias}. In this article we
propose a new group and a subgroup of the group of automorphisms for the MOR
cryptosystem. Our group is the \emph{group of unitriangular matrices
  over a finite field} and the automorphisms are the composition of
\emph{diagonal, inner and central automorphisms}. We show that for this
group and subgroup of automorphisms, MOR is as secure as the
El-Gamal cryptosystem over finite fields.

There is still a lot of interest in cryptosystems using the
discrete logarithm problem in finite fields, for example, the El-Gamal
cryptosystem. We claim that we had a reasonable amount of success with
these groups and automorphisms. Though the most desirable consequence
of this research would be no \emph{sub-exponential} attack on the
cryptosystem.  

There is one other shift in our proposed MOR cryptosystem. We
are using \emph{polycyclic} groups \cite[Chapter 9]{sims} for the
cryptosystem; computation 
with this class of groups is done differently than with the
multiplicative group of finite fields. We are yet to understand the
consequence of this shift, from arithmetic in finite fields to
arithmetic in a polycyclic group and the use of automorphisms instead
of exponentiation. 

It is often expected of the proposer of a new cryptosystem to provide
parameters and to show that the cryptosystem is \emph{semantically
  secure}\footnote{For our definition of semantic security see
  \cite{crypto98}. Briefly stated, a
  cryptosystem is semantically secure if it is secure against a
  passive eavesdropper.}. The El-Gamal encryption scheme is considered
semantically 
secure \cite{crypto98} and so it remains to be seen if the
proposed MOR cryptosystem is also
semantically secure. Note that the semantic security of the MOR
cryptosystem depends on the group used \cite[Section 3]{tobias}.

We are not yet in a position to provide parameters because
the discrete logarithm problem in the automorphism group, on which the
security of our cryptosystem depends, is not well studied. Moreover,
since the best known attack on the proposed MOR cryptosystem is the
discrete logarithm problem in 
finite fields, hence one can pick parameters from any cryptosystem
using the discrete logarithm problem, e.g., the El-Gamal
cryptosystem and use it for the proposed MOR cryptosystem. The MOR
cryptosystem is a straightforward generalization of 
the El-Gamal cryptosystem, so it is easy to see that MOR is not
secure against \emph{indistinguishability-secure from
  chosen-ciphertext attack} \cite[Section 2]{menezes-koblitz},
however ideas similar to the Cramer-Shoup cryptosystem
\cite{crypto98} should make it achieve any security goal in any attack
model.     
\section{The MOR cryptosystem}
In this section we discuss the MOR cryptosystem \cite{crypto2001} and
critique some of the points discussed by the authors. There are
two different security concepts used in \cite{crypto2001}.
\begin{description}
\item[i.] The discrete logarithm problem in the group of inner automorphisms.
\item[ii.] Membership problem in a finite cyclic group.
\end{description}
Let us describe the MOR cryptosystem in details. Let $G=\langle\gamma_1,\gamma_2,\ldots,\gamma_s\rangle$ be a finite
non-abelian group. Let $\phi_g$ be an inner automorphism of $G$
defined by $\phi_g(x)=g^{-1}xg$ for all $x\in G$. Then
$\phi_g^m(x)=g^{-m}xg^m$ for all $x\in G$ and $m$ a
positive integer. We are working in the group of inner automorphisms
with the composition of automorphism as the group operation.
Now suppose Eve wants to set up a public key for
herself. Then she chooses $g$ and publishes $\phi_g$ and
$\phi_g^m$. She, however, doesn't publish $g$ and $g^m$; instead she
publishes $\{\phi_g(\gamma_i)\}_{i=1}^s$ and
$\{\phi_g^m(\gamma_i)\}_{i=1}^s$. Then to send a message (plaintext)
$a\in G$, Bob computes $\phi_g^r$ and $\phi_g^{mr}$ from the
public information, for a random $r\in\mathbb{N}$ and then computes
$\phi_g^{mr}(a)$. He then sends Eve 
$\left(\phi_g^r,\phi_g^{mr}(a)\right)$. As in the El-Gamal cryptosystem
Alice, knowing $m$, can compute $\phi_g^{mr}$ from $\phi_g^r$ and, hence, the inverse $\phi_g^{-mr}$ and the plaintext $a$.

What does the security of this protocol depend on? Firstly, if one can
solve the discrete logarithm problem in $\phi_g$ and $\phi_g^m$ then
the protocol is broken. On the other hand, since the inner
automorphisms are presented as the action on generators, it might be
difficult to find $g$ from the public information 
$\{\phi_g(\gamma_i)\}_{i=1}^s$. Moreover, $\phi_g=\phi_{gz}$ for any
$z\in Z(G)$ the center of the group $G$, so even if there is an
algorithm to find $g$, that $g$ 
might not be unique. The authors of the MOR cryptosystem uses this
fact for security as follows: suppose one knows the $g$ from
$\phi_g$ and then tries to determine the $g^m$ in $\phi_{g^m}$ then by
solving the conjugacy problem they will come up with $g^mz$. Then they will  
have to solve the membership problem in the cyclic group $\langle
g\rangle$ before they 
can even try to solve the discrete logarithm problem. Of course this
attack on the system does not include that someone might be able to
solve for $m$ from the public informations
$\{\phi_g(\gamma_i)\}_{i=1}^s$ and
$\{\phi_{g^m}(\gamma_i)\}_{i=1}^s$. Moreover, as shown in \cite[Theorem
1]{asiacrypt2004} there is an effective way using only \emph{black box group}
operations to get around this 
membership problem by switching to the discrete logarithm problem in $G/Z(G)$.

The idea behind this scheme seems to be novel and the idea of using the
membership problem in public key cryptography might have interesting
applications. However, the biggest test for an idea to develop a public
key protocol is the ability to find groups
that produce fast encryption, fast decryption and is secure.  

The idea of using automorphisms; where the public information about
these automorphisms is its action on generators puts severe
restrictions on the groups useful in this scheme.
\begin{description}
\item The groups used should have a fast algorithm to express an
  element as a word in generators. Unless every group
  element is presented as words in generators, e.g., polycyclic
  groups where fast collection algorithms are available, this is hard
  to achieve.
\end{description}
What concerns us the most is the use of two different
cryptographic primitives -- the discrete logarithm problem and the
membership problem simultaneously! It can be argued that two
insecure locks do not make one secure lock; just get two different
person to work 
on them simultaneously or use a \textit{meet in the middle}
attack. The converse of 
the idea is that one secure lock is enough to guard a secret. Stated
plainly, the idea of using the membership problem and the discrete
logarithm problem simultaneously in a protocol is probably not
wise. On top of this, since MOR is a generalization of the El-Gamal
cryptosystem whose security depends on the discrete logarithm problem, the
computational Diffie-Hellman problem and the decision Diffie-Hellman
problem \cite[Section 2.3]{thesis}or \cite[Section
2]{menezes-koblitz}; this cryptosystem is not ideally 
suited to exploit the 
membership problem. This was echoed in \cite{paeng}. In the
definition of the MOR cryptosystem in \cite{paeng} the whole automorphism group
was considered instead of the group of inner automorphisms as in
\cite{crypto2001}, and the 
requirement that the automorphisms be presented as action on generators
was dropped. Following that: in this article we won't use the membership
problem; we will rely on the discrete logarithm problem in the
automorphism group for security. 

The basic scheme for a MOR cryptosystem is as follows and is an
adaptation of \cite[Section 2]{paeng}:

Let $G$ be a group and $\phi:G\rightarrow G$ be an
automorphism. In this paper, if we work with automorphisms of
$G$, we work in the automorphism group of $G$, with the group
operation being the composition of automorphisms. 
\subsection{Description of the MOR cryptosystem}\label{MOR}
\noindent Alice's keys are as follows:
\begin{description}\label{keyex}
\item[Public Key] $\phi$ and $\phi^m$, $m\in\mathbb{N}$.
\item[Private Key] $m$.
\end{description}
\paragraph{\textbf{Encryption}}
\begin{description}
\item[a] To send a message $a\in G$ Bob computes $\phi^r$ and $\phi^{mr}$ for a
random $r\in\mathbb{N}$.
\item[b] The ciphertext is $\left(\phi^r,\phi^{mr}(a)\right)$.
\end{description}
\paragraph{\textbf{Decryption}}
\begin{description}
\item[a] Alice knows $m$, so if she receives the ciphertext
  $\left(\phi^r,\phi^{mr}(a)\right)$, she computes $\phi^{mr}$ from $\phi^r$ and
  then $\phi^{-mr}$ and then from $\phi^{mr}(a)$ computes $a$.
\end{description}
Alice can compute $\phi^{-mr}$ two ways; if she has the information
necessary to find out the order of the automorphism $\phi$ then she can use
the identity $\phi^{t-1}=\phi^{-1}$ whenever $\phi^t=1$. Also, she can
find out the order of some subgroup in which $\phi$ belongs and
use the same identity. However, the smaller the subgroup, more efficient
the decryption algorithm.
\section{Proposed group for the MOR cryptosystem}\label{group}
The non-abelian group we are proposing for the MOR cryptosystem is the group of
unitriangular matrices over a finite field $\mathbb{F}_q$ of
characteristic $p$, where $p$ is a
prime number. The group of unitriangular
matrices over $\mathbb{F}_q$ is often denoted by
$UT(n,q)$. This group consists of all square 
matrices of dimension $n$; the diagonal elements are $1$ (the
multiplicative identity of the field) and all entries below the
diagonal are $0$ (the additive identity of the field). The entries
above the diagonal can be any element of the finite field $\mathbb{F}_q$. The
group operation is matrix multiplication. An
arbitrary element $g\in UT(4,q)$ looks like,
\[g=\left(\begin{array}{cccc}
1 & \ast &\ast &\ast\\
0 & 1     &\ast &\ast\\
0 & 0     &1     &\ast\\
0 & 0     &0     &1
\end{array}\right).\] The $\ast$ denotes a field
element. From a simple counting argument it follows that $UT(n,q)$ is
a Sylow $p$-subgroup of the general linear group $GL(n,q)$ where $p$
is the characteristic of the finite field $\mathbb{F}_q$.

Let $e_{ij}$ for $i<j$ represent the matrix with $1$ in the $(i,j)$
position and $0$ elsewhere. It is customary to represent $g\in
UT(n,q)$ as $1+\sum\limits_{i<j}a_{ij}e_{ij}$, where $a_{ij}\in
\mathbb{F}_q$. Notice that $1$ above is the identity matrix. We will
abuse the notation a little bit and use $1$ as the identity of
$UT(n,q)$ and $\mathbb{F}_q$ simultaneously. It should be
clear from the context which $1$ we are referring to.

There are two fundamental set of relations in
$UT(n,q)$ along with the relations in the field
$\mathbb{F}_q$. For $(1+ae_{ij}),\;(1+be_{kj})\in UT(n,q)$
where $a,b\in\mathbb{F}_q$ they are as follows: 

\begin{equation}\label{eqn1}(1+ae_{ij})(1+be_{ij})=1+(a+b)e_{ij}\end{equation}
\begin{equation}\label{eqn2}[1+ae_{ij},1+be_{kl}]=
\left\{
\begin{array}{ccc}
1+abe_{il} &\text{if}&j=k,\;\; i\neq l\\
1-abe_{kj} &\text{if}&i=l,\;\; j\neq k\\ 
1          &\text{otherwise}&
\end{array}\right.\end{equation}
Here $[x,y]=x^{-1}y^{-1}xy$ is the commutator of elements $x,y\in G$ for any
group $G$. It is well known that the additive group of $\mathbb{F}_q$,
often written as $\mathbb{F}_q^+$,
is a $\gamma$ dimensional vector space over $\mathbb{Z}_p$, where $p^\gamma=q$.  
It follows \cite[Page 455]{weir} that the minimal set of generators
of $UT(n,q)$ are $1+\delta_ke_{i,i+1}$,
$k=1,2,\ldots,\gamma$ and $i=1,2,\ldots,n-1$. The set
$\{\delta_1,\delta_2,\ldots,\delta_\gamma\}$ is a basis
of $\mathbb{F}_q^+$ over $\mathbb{Z}_p$. The center of
$UT(n,q)$ is $1+ke_{1,n}$ 
where $k\in\mathbb{F}_q$.

Since $UT(n,q)$ is a finite p-group, it is a finite
\emph{nilpotent} group and a \emph{polycyclic} group \cite[Proposition
3.4]{sims}. 
\begin{definition}[Polycyclic Group]
A group $G$ is a polycyclic group if there is a finite chain of
subgroups $G=G_1\supset G_2\supset\ldots\supset
G_k\supset G_{k+1}=1$ such that $G_{i+1}$ is a normal subgroup of
$G_i$ and $G_i/G_{i+1}$ is cyclic.
\end{definition}
Since in a polycyclic group $G$, $G_i/G_{i+1}$ is cyclic, there
is an $a_i$ in $G_i$ such that the image of $a_i$ in $G_i/G_{i+1}$
generates $G_i/G_{i+1}$. It is easy to see that
$\{a_1,a_2,\ldots,a_k\}$ generates the group $G$ and is known as
the \emph{polycyclic generating set}. Since we are dealing with finite
groups, $|G_{i+1}:G_i|=m_i$ is finite. It follows that (see
\cite[Section 9.4]{sims}) every word in $G$ can be expressed uniquely
as $a_1^{\alpha_1}a_2^{\alpha_2}\ldots
a_k^{\alpha_k}$ where $0\leq \alpha_j <m_j$ for
$j=1,2,\ldots,k$. These words are called \emph{collected words}. Using a
collection algorithm \cite[Section 9.4]{sims} any word in
$\{a_1,\ldots,a_k\}$ can be expressed as a collected word. So, in
this group computing the inverse and the product is fast and easy,
i.e., there is a fast implementation of polycyclic groups and their
arithmetic \cite[Polycyclic Package]{GAP4}. 

Let us talk about a polycyclic generating set of
$UT(n,p)$; for an arbitrary finite field $\mathbb{F}_q$ this can be
similarly done. For sake of simplicity we take $n=4$. Let
$a_1=1+e_{12}$, $a_2=1+e_{23}$, $a_3=1+e_{34}$, $a_4=1+e_{13}$,
$a_5=1+e_{24}$ and $a_6=1+e_{14}$. It is shown in \cite[Section
9.4, Example 4.1]{sims} that $\{a_1,a_2,\ldots,a_6\}$ forms a polycyclic
generating set for $UT(4,\mathbb{Z})$. It is easy to see that this is
also a polycyclic generating set for $UT(4,p)$ for an arbitrary
prime $p$. The polycyclic generating set for $UT(n,p)$ can be
similarly found for an arbitrary $n$.
\subsection{The diagonal automorphism}\label{diagonal}
Let $D$ be an diagonal matrix, i.e., a matrix of dimension $n$ over the
field $\mathbb{F}_q$, and the only non-zero elements are in the diagonals. We
will represent a diagonal matrix $D$ as $[w_1,w_2,w_3,\ldots,w_n]$, where
$w_i$ are non-zero elements of the field $K$ and are the diagonal
elements of the matrix $D$. It is easy to see that if
$w_1=w_2=\ldots=w_n$ then the diagonal matrix is a scalar
matrix. Weir\cite[Section 4]{weir} introduced the 
diagonal automorphisms on $UT(n,q)$. Let $D$ be a diagonal matrix
given by $[w_1,w_2,\ldots,w_n]$; then from matrix multiplication it
follows that
$D^{-1}xD$ for an $x\in UT(n,q)$ where
$x=1+\sum\limits_{i<j}a_{ij}e_{ij}$ is given by
$1+\sum\limits_{i<j}(w_i^{-1}a_{ij}w_j)e_{ij}$. Since the scalar matrices have
the same diagonal elements, the group of diagonal automorphisms
has order $(q-1)^{n-1}$.    

These diagonal automorphisms are not inner automorphisms
because the diagonal matrices are not unitriangular. We will now study
the MOR cryptosystem using these diagonal automorphisms. It is easy to
see that if $D=[w_1,w_2,\ldots,w_n]$ and $\phi(x)=D^{-1}xD$ for $x\in
UT(n,q)$ then $\phi^m(x)=D^{-m}xD^m$ where
$D^m=[w_1^m,w_2^m,\ldots,w_n^m]$ where $m\in\mathbb{N}$. So, if Alice
makes $D$ and $D^m$ public then finding the $m$ is solving the discrete
logarithm problem in the multiplicative group $\mathbb{F}_q^\times$ of
the finite field $\mathbb{F}_q$.

If the plaintext is $a\in UT(n,q)$, then computing
$\phi^m(a)$ is easy and can be done easily from the formula above. So,
using these diagonal 
automorphisms one can have a secure protocol similar to that of the
El-Gamal cryptosystem. Clearly, there is no advantage for using this
protocol over 
El-Gamal; the security depends on the discrete logarithm problem in
the multiplicative group of the 
finite fields; but one has to do more work than the El-Gamal
cryptosystem for encryption and decryption. 

If we take the group $UT(2,q)$ of $2\times 2$ unitriangular matrix over the
finite field 
$\mathbb{F}_q$, then for a $x\in\mathbb{F}_q^{\times}$ we can consider
a diagonal 
automorphism presented on the generator of this group as 
\[\phi:=
\begin{pmatrix}
1&1\\
0&1
\end{pmatrix}
\mapsto
\begin{pmatrix}
1&x\\
0&1
\end{pmatrix}
\;\;\text{and the $m^{\text{th}}$ power} 
\;\;\phi^m:=\begin{pmatrix}
1&1\\
0&1
\end{pmatrix}
\mapsto
\begin{pmatrix}
1&x^m\\
0&1
\end{pmatrix}.
\]
If we use the MOR protocol as in Section \ref{MOR} with these
automorphisms, then it is identical to 
the El-Gamal cryptosystem over a finite field. 

So, we claim that the MOR cryptosystem as in Section \ref{MOR} with the
diagonal automorphisms is computationally and semantically secure and
can be made 
indistinguishability-secure from chosen-ciphertext attack
using ideas similar to the Cramer-Shoup
cryptosystem\cite{crypto98}. Notice that it is essential for 
the above mentioned use, that the $w_i$ are all different from one
another; otherwise valuable information about the plaintext will be leaked.
\subsection{The inner automorphism} \label{inner}
Inner automorphisms are the
easiest of the automorphisms to study; they are defined as
$I_g(x)=g^{-1}xg$ for all $x\in UT(n,q)$ and $g\in
UT(n,q)$. It is 
well known that the group of inner automorphisms $I(G)$ for an
arbitrary group $G$ is a normal subgroup of the automorphism group of
$G$. It is also known 
that $I(G)$ is isomorphic to $G/Z(G)$. From which it follows that the
order of the group of inner automorphisms of the group $UT(n,q)$ is $q^{\frac{n^2-n-2}{2}}$.
We will now see what happens if
we use the inner automorphisms for the MOR cryptosystem.

Let $\phi=I_g$ as described in the MOR cryptosystem (see Section \ref{MOR}). Since the conjugacy problem is easy
and we are not using the membership problem, we can safely assume that
$g$ and $g^m$ is public. If 
\[g=\left(\begin{array}{cccc}
1 & a_{12}&a_{13}&a_{14}\\
0&1&a_{23}&a_{24}\\
0&0&1&a_{34}\\
0&0&0&1
\end{array}\right)\]
then 
\[g^m=\left(\begin{array}{cccc}
1 & ma_{12}&\ast&\ast\\
0&1&ma_{23}&\ast\\
0&0&1&ma_{34}\\
0&0&0&1
\end{array}\right)\] where $\ast$ represents a field element.

Now the discrete logarithm problem to find $m$ essentially becomes
the discrete logarithm problem in 
$\mathbb{F}_q^+$. Since the discrete logarithm problem in the additive
group of a 
finite field is known to be easy, we do not believe that using only
inner automorphisms one can build a secure MOR cryptosystem.
\subsection{The central automorphism}\label{central}
The group of central automorphisms is the group most widely studied
after the group of inner automorphisms. The reason of its popularity
is that the group of central automorphisms is the group of centralizers of the
group of inner automorphisms, i.e., the central automorphisms commute
with the inner automorphisms and fix the derived subgroup
elementwise. It can be shown that if $\psi$ is a 
central automorphism of a group $G$ then $\psi(g)=gz_g$ where $z_g\in
Z(G)$ and depends on $g$. It follows \cite{levchuk} that a description of the
central automorphism $\zeta_r(\lambda)$ of $UT(n,q)$ is 
\[\zeta_r(\lambda):1+a_{r,r+1}e_{r,r+1}\mapsto
1+a_{r,r+1}e_{r,r+1}+\lambda\left(a_{r,r+1}\right)e_{1,n}\]where
$\lambda$ is an endomorphism of $\mathbb{F}_q^+$ and
$r=1,2,\ldots,n-1$. Now since $\lambda$ 
is an endomorphism and $\mathbb{F}_q^+$ is a $\gamma$-dimensional
vector space over $\mathbb{Z}_p$, if $\lambda(\delta_i)=b_i$ for
$i=1,2,\ldots,\gamma$ then   
we arrive at \cite[Page 463]{weir} where
a description of the central automorphisms for the $UT(n,q)$ is given
as $1+\delta_ie_{r,r+1}\mapsto 1+\delta_ie_{r,r+1}+b_ie_{1,n}$ where 
$r=1,2,\ldots,n-1$, $b_i$ is an arbitrary element of $\mathbb{F}_q$. This can
also be represented as $1+\delta_ie_{r,r+1}\mapsto
(1+\delta_ie_{r,r+1})(1+b_ie_{1,n})$. So composing this map $n$ times gives
us $1+\delta_ie_{r,r+1}\mapsto (1+\delta_ie_{r,r+1})(1+nb_ie_{1,n})$. Notice
that if $r=1,n-1$ then the central automorphisms are inner
automorphisms and from this it follows that the order of the group of
central automorphisms is $q^{\gamma(n-3)}$ where $p^\gamma=q$ (see
\cite[Page 463]{weir}).
Since the description of the central automorphisms depend on $\lambda$,
unlike the inner or the diagonal automorphisms the only possible
description of a central automorphism is by action on generators of
the group $G$. 

So, if we
take a central automorphism to use in the MOR cryptosystem then from the
public information the discrete logarithm problem is the same as the
discrete logarithm problem in $\mathbb{F}_q^+$. The discrete
logarithm problem in the additive group of a finite 
field is easy; central automorphisms alone do not provide us with a
secure MOR cryptosystem.   
\section{A proposed automorphism for the MOR cryptosystem}
Currently the proposed group for the MOR cryptosystem
\cite{crypto2001} is $SL(2,\mathbb{Z}_p)\rtimes\mathbb{Z}_p$. This is
a split extension of $SL(2,\mathbb{Z}_p)$ by $\mathbb{Z}_p$. The
automorphisms proposed are the inner automorphisms. It is
shown in \cite[Theorem 2]{paeng} that the discrete logarithm problem
in the group of inner automorphisms of
$SL(2,\mathbb{Z}_p)\rtimes\mathbb{Z}_p$ is the same as the discrete
logarithm problem in $SL(2,\mathbb{Z}_p)$. In \cite{menezes1} the authors
show that the discrete logarithm problem in $GL(n,q)$, the general
linear group over the finite field $\mathbb{F}_q$, is at most as hard
as the discrete logarithm problem in some finite extension field of
$\mathbb{F}_q$. Since there are sub-exponential attacks on the discrete
logarithm problem in finite fields such as the index calculus attack,
there is every reason (practical as well as academic) to look for non-abelian
groups and automorphisms in these groups in search for a better MOR
cryptosystem.  

In \cite{asiacrypt2004} the authors developed a
\emph{central commutator attack}; they showed that inner automorphisms
are not well suited for MOR cryptosystem; especially when the group is
nilpotent. 

So, it is now clear that if we are using nilpotent groups, ($UT(n,q)$ is a
finite p-group and hence nilpotent) then we have to look for outer
automorphisms. The diagonal and the central automorphisms are outer
automorphisms. On the other hand, as we saw in the last section,
diagonal automorphisms do provide us with a secure MOR cryptosystem
and the only way to represent a central automorphism is its action on
generators. The security with diagonal automorphisms turns out
to be the discrete logarithm problem in the 
multiplicative group of the finite field, and the
central and the inner automorphisms from their presentation reveals
valuable information. 

Now we are in a position to describe and justify the automorphism
group that we are going to propose for the MOR cryptosystem, it is 
\begin{quote}
\centerline{central \emph{composed} inner \emph{composed} diagonal
  automorphism.}\end{quote} 
Let us denote by $\mathcal{I}$, $\mathcal{D}$ and $\mathcal{L}$ the
group of inner, diagonal and the central automorphisms of $UT(n,q)$
respectively. It is well known that the centralizer of a normal
subgroup in a group $G$ is normal in $G$. The subgroup $\mathcal{I}$ is
normal in the automorphism group of $UT(n,q)$ and so is
$\mathcal{L}$. So,
$\mathcal{I}\mathcal{L}$ is a subgroup of the automorphism group
of $UT(n,q)$. The diagonal automorphisms do not commute with the
inner automorphisms, the
group of automorphisms we plan on using are elements of the subgroup
$(\mathcal{I}\mathcal{L})\rtimes\mathcal{D}$. It clearly
follows that the subgroup of the above automorphisms
have order $$q^{\frac{n^2-n-2}{2}}\times(q-1)^{n-1}\times
q^{\gamma(n-3)}\;\;\text{where}\;\; p^{\gamma}=q.$$

We saw earlier that the discrete logarithm problem in the group of
diagonal automorphisms is at most as secure as the discrete logarithm
problem in the finite field. 

We were hoping that by composing a diagonal automorphism
with the inner and central automorphism we might be able to diffuse
the public information, so that, the reduction to the discrete
logarithm problem in the finite field becomes impossible. We now show
by means of a small 
example that with the best of 
efforts we are not able to beat the sub-exponential attack on finite
fields. 
\subsection{A small example}
We now explain the MOR cryptosystem with a small example. We used
\cite[Polycyclic Package]{GAP4} for this example, notations are from
Section \ref{group}. We choose $n=4$ and
$q=1297$ where $1297$ is a prime. We pick three random integers
$984$, $807$ and $452$. Then we define a central
automorphisms (see Section \ref{central}) $map1$ as \[map1=\left\{
\begin{array}{cc}
a_1\longrightarrow a_1a_6^{984}\\
a_2\longrightarrow a_2a_6^{807}\\
a_3\longrightarrow a_3a_6^{452}
\end{array}\right.\]all other generators remain fixed. Note that a
central automorphism fixes commutators.
Next we pick a random element
$h:=a_1^{83}a_2^{462}a_3^{1202}a_4^{1209}a_5^{793}a_6^{152}$ and
compute the inner automorphism (see Section \ref{inner}), $map2:\;\;x\mapsto h^{-1}xh$ corresponding to
$h$. 
\[map2:=\left\{
\begin{array}{ccc}
a_1&\longrightarrow & a_1a_4^{462}a_6^{1001}\\
a_2&\longrightarrow & a_2a_4^{1214}a_5^{1202}a_6^{103}\\
a_3&\longrightarrow & a_3a_5^{835}a_6^{88}\\
a_4&\longrightarrow & a_4a_6^{1202}\\
a_5&\longrightarrow & a_5a_6^{1214}\\
a_6&\longrightarrow & a_6
\end{array}\right.\]
Then we take the diagonal automorphism (see Section \ref{diagonal})
corresponding to 
$[624,155,538,126]$, the diagonal automorphism $map3$ is
\[map3=\left\{
\begin{array}{cc}
a_1&\longrightarrow a_1^{576}\\
a_2&\longrightarrow a_2^{1267}\\
a_3&\longrightarrow a_3^{574}\\
a_4&\longrightarrow a_4^{878}\\
a_5&\longrightarrow a_5^{938}\\
a_6&\longrightarrow a_6^{736}
\end{array}\right.\]
Then the automorphism Alice will make public is
$\phi=map1\cdot map2\cdot map3$ and that is given by
\[\phi=\left\{
\begin{array}{ccc}
a_1&\longrightarrow &a_1^{576}a_4^{972}a_6^{538}\\
a_2&\longrightarrow &a_2^{1267}a_4^{1055}a_5^{383}a_6^{508}\\
a_3&\longrightarrow &a_3^{574}a_5^{1139}a_6^{558}\\
a_4&\longrightarrow &a_4^{878}a_6^{118}\\
a_5&\longrightarrow &a_5^{938}a_6^{1168}\\
a_6&\longrightarrow &a_6^{736}
\end{array}\right.\]
and if Alice chooses her private key to be $65$ then
\[\phi^{65}=\left\{
\begin{array}{ccc}
a_1&\longrightarrow & a_1^{450}a_4^{1145}a_6^{618}\\
a_2&\longrightarrow & a_2^{1263}a_4^{1269}a_5^{1242}a_6^{1093}\\
a_3&\longrightarrow & a_3^{526}a_5^{708}a_6^{279}\\
a_4&\longrightarrow & a_4^{264}a_6^{1190}\\
a_5&\longrightarrow & a_5^{274}a_6^{836}\\
a_6&\longrightarrow & a_6^{85}
\end{array}\right.\]
The automorphisms $\phi$ and $\phi^{65}$ are public, (see description of
the MOR cryptosystem in Section \ref{keyex}). Notice
that $(576)^{65}\mod 1297=450$. An observant reader will further notice that 
from the public information of $\phi$ and $\phi^{65}$ 
that if $k_j^\prime$ is the exponent of $a_j$ in
$\phi^{65}(a_j)$ and if $k_j$ is the exponent of $a_j$ in $\phi(a_j)$
for $j=1,2,3$ and $j=6$, then $k_j^\prime$ is $k_j^{65}$. The
reason for this is that the inner and the central automorphisms leave
the exponent of
$a_1,a_2,a_3,a_6$ unchanged in the image as seen in map1 and map2. The
only thing that changes 
$\{a_1,a_2,a_3,a_6\}$ is the diagonal automorphism and then the change
is $a_j\mapsto a_j^{w_j^{-1}w_{j+1}}$ for $j=1,2,3$ and
$a_6\mapsto a_6^{w_1^{-1}w_4}$. Then composing the map $m$ times
gives us $a_j\mapsto a_j^{\left({w_j^{-1}w_{j+1}}\right)^m}$ for $j=1,2,3$ and
$a_6\mapsto a_6^{\left({w_1^{-1}w_4}\right)^m}$. 
 
This leads us to the best known attack against this cryptosystem. If
one can solve the discrete logarithm problem in a finite field then he
can figure out the $m$ from the public information of $\phi$ and
$\phi^m$ as demonstrated above. There are sub-exponential algorithms,
such as the index calculus methods, in finite fields to solve the discrete
logarithm problem.

\section{The security of the proposed MOR cryptosystem}
If we assume that MOR using $UT(n,q)$ with proposed automorphisms is
broken for an arbitrary $n$, then it is broken in $UT(2,q)$ with
diagonal automorphisms. The MOR cryptosystem using $UT(2,q)$ is similar
to the El-Gamal cryptosystem over finite fields (see Section \ref{diagonal}).
This breaks the El-Gamal cryptosystem over finite fields. Conversely,
if the El-Gamal 
cryptosystem over finite fields is broken by solving DLP in finite
fields then one can break the proposed MOR cryptosystem. This is clear
from the action of the automorphisms on the elements as described
before and is also clear from the example above. So, we claim that in
terms of security, the proposed MOR cryptosystem is equivalent to the
El-Gamal cryptosystem over finite fields.
\section{Conclusion}
In this paper we studied a new non-abelian finite group and a group of
outer automorphisms for the MOR cryptosystem. The computational
security of any proposed 
cryptosystem is always an open question.
This is the first time that the group of
unitriangular matrices and automorphisms over it has been proposed for public
key cryptography; more work needs to be done to assure one of the
security of the said system.

This article clearly shows that the MOR cryptosystem has a lot to
offer to the public key cryptography. We showed that with
the right kind of groups, the MOR cryptosystem can offer a secure
cryptosystem.

\textbf{Acknowledgements:} This paper was written when the author was
visiting the Applied Statistics Unit of the Indian Statistical Institute
at Kolkata. The author expresses his gratitude to Bimal Roy for
making this visit possible. The author received help from Bettina Eick
regarding computations with GAP\cite{GAP4} which he greatfully acknowledges.
\nocite{*}
\bibliography{paper}
\bibliographystyle{amsplain} 
\end{document}